\documentclass[12pt,prd,amsfonts,nofootinbib,showpacs]{revtex4}
\usepackage{graphicx, epsfig}
\newcommand{\be}{\begin{equation}}
\newcommand{\ee}{\end{equation}}
\newcommand{\bea}{\begin{eqnarray}}
\newcommand{\eea}{\end{eqnarray}}
\newcommand{\nn}{\nonumber}
\newcommand{\x}{\times}
\begin{document}
\title{$\Delta L=2$ hyperon semileptonic decays}

\author{C. Barbero$^1$, Ling-Fong Li$^2$, G. L\'opez Castro$^3$ and A.
Mariano$^1$}
\affiliation{$^1$ Departamento de F\'\i sica, Universidad Nacional de La
Plata, cc 67, 1900
La Plata, Argentina \\ 
$^2$ Department of Physics, Carnegie Mellon University, Pittsburg, PA
15213, USA \\ 
$^3$Departamento de F\'\i sica, Cinvestav, Apartado
Postal 14-740, 07000 M\'exico D.F., M\'exico}
\begin{abstract}
We compute the rates of semileptonic $B_A \to B_Bl^-l^-$ ($l=e$ or $\mu$)
hyperon transitions in a model where intermediate states involve loops
of baryons and a Majorana neutrino. These rates turn out to be
well below present experimental bounds and other theoretical estimates.
From the experimental upper limit on the $\Xi^- \to p\mu^-\mu^-$ decay,  
we derive the bound $\langle m_{\mu \mu} \rangle \leq 22$ TeV for
the effective Majorana mass of the muon neutrino. Also, an estimate of 
background contributions for these decays due to the allowed $B_A \to
B_Bl^-l^-\bar{\nu}\bar{\nu}$ decays are provided.
 \end{abstract}
\maketitle


\section{Introduction}

   Non-degenerated neutrino masses provide at present the most accepted 
explanation for the well established experimental results on neutrino
oscillations \cite{oscillations}. Nowadays, strong experimental and
theoretical efforts are focused on trying to determine the absolute
values of neutrinos masses \cite{Elliott:2002xe}. Of particular interest
in this regard is the question about whether neutrinos are Dirac or
Majorana particles. A crucial role to address this question is being
played by several experiments looking to the possible existence of
$|\Delta L| =2$ transitions \cite{Elliott:2002xe}.

   In this paper we focus on the $\Delta L=2$, $\Delta S=1$ and $2$   
transitions between spin-1/2 hyperons, $B_A \to B_B l^-l^-$, following a
procedure discussed before in Ref. \cite{Barbero:2002wm} in the
case of the $\Delta S=0$ 
 $\Sigma^- \to \Sigma^+e^-e^-$ decays. Up to now, little
attention has been paid to these decays because searches on neutrinoless
double beta decays of nuclei are far more sensitive probes to effects of
Majorana electron neutrinos.
 Eventhough experiments on hyperon decays can not reach very small
branching fractions as in nuclear decays, it is worth to mention that
there
channels involving muons which are not available in nuclear decays. 

Nevertheless, a few experimental bounds on hyperon $|\Delta L|=2$
transitions have become available recently \cite{pdg}. From a
data analysis of an old BNL experiment, authors of Ref.
\cite{Littenberg:1991rd} have derived $B(\Xi^- \to p\mu^-\mu^-) < 3.7
\times 10^{-4}$ at 90$\%$ c.l.. More recently, the HyperCP Collaboration
has reported an improved bound $B(\Xi^- \to p\mu^-\mu^-) < 4.0
\times 10^{-8}$ at 90$\%$ c.l. on this decay mode \cite{Rajaram:2005bs}.
Besides this, the other experimental bound reported
so far is the $\Delta L=-2$ decay mode of the charmed baryon
$\Lambda^+_c$ with a branching fraction of   $B(\Lambda_c^+ \to
\Sigma^-\mu^+\mu^+) < 7 \times 10^{-4}$ also at 90$\%$ c.l. 
\cite{Kodama:1995ia}. The above decays can be useful in bounding the
effective Majorana mass of the muon neutrino, which (rather poor)
present limit $\langle m_{\mu\mu} \rangle < 0.04$ TeV
\cite{Littenberg:2000fg} comes
from an indirect bound on the $K^+ \to \pi^-\mu^+\mu^+$ decay
\cite{Appel:2000tc}. 

  On the theoretical side, the only available studies about $\Delta
L=2$ hyperon decays have been reported in Refs. \cite{Barbero:2002wm,
Li:2007qx}. Based on the dynamics of weak interactions for these
processes and phase-space considerations, Ref. \cite{Li:2007qx} suggests 
that a branching ratio of about $10^{-10}$ may be expected for such
decays in one of the most optimistic new physics scenarios. Just for
comparison, let us mention that  an explicit calculation done in Ref.
\cite{Barbero:2002wm} for the $\Delta S=0$ hyperon decay gives
$B(\Sigma^- \to
\Sigma^+e^-e^-) \approx 1.49 \times 10^{-35}$ for an effective
electron neutrino mass of about $\langle m_{\nu e} \rangle = 10$ eV. Thus,
we may expect a large suppression of $\Delta L=2$ decay rates in a light
Majorana neutrino scenario and it is the purpose of our paper to explore
in further detail this possibility through an explicit calculation.

\section{Hyperon $\Delta L=2$ decays}

  In this paper we will consider the $\Delta L=2$ hyperon decays listed in
Table
I. We use a model where the dominant contributions are given by
loops involving  virtual baryon and Majorana neutrino states (see Figure
1) \cite{Barbero:2002wm}. The properly antisymmetrized decay amplitude for
this process is given
by:
\be
{\cal M}_{0\nu}=G^2({\cal M}_1 -{\cal M}_2)\ ,
\label{1}
\ee
where $G^2$ is the effective weak coupling ($G_F$ is the Fermi constant, 
and $V_{ij}$ the $ij$ entry of the Cabibbo-Kobayashi-Maskawa matrix):
\be
G^2=G_F^2\times \ \left\{ \begin{array}{c}
V_{ud}^2\ \ \ \ \ \   {\rm for }\ \Delta S=0, \\
V_{ud}V_{us}\ {\rm for } \  \Delta S= 1, \\
V_{us}^2\ \ \ \  {\rm for }\ \Delta S= 2\ . \end{array} \right.
\label{2}
\ee

\begin{table}
\begin{tabular}{|c|c|c|}
\hline
$\Delta S=0$ & $\Delta S =1$ & $\Delta S=2$ \\
\hline
$\Sigma^-\to \Sigma^+e^-e^-$ & $\Sigma^- \to  pe^-e^-$ & $\Xi^- \to
pe^-e^-$ \\
 & $\Sigma^- \to p\mu^-\mu^-$ & $\Xi^-\to p\mu^-\mu^-$ \\
 & $\Xi^-\to \Sigma^+e^-e^-$ & \\
\hline   
\end{tabular}
\caption{$\Delta L=2$ modes of spin-1/2 hyperon semileptonic transitions.}
\end{table}

The expressions for the decay amplitudes defined in Eq. (\ref{1}) are
\cite{Barbero:2002wm} ($i=1,\ 2$):
\be
{\cal M}_i=\sum_{j}m_{\nu j}U_{lj}^2\int
\frac{d^4q}{(2\pi)^4}
\frac{1}{q^2-m_{\nu 
 j}^2}L^{\alpha\beta}(p_i,p_{3-i})H_{\alpha\beta}(Q_i(q))\ ,
\label{17}
\ee
where $Q_i(q)\equiv p_B+p_i-q=p_A-p_{3-i}-q$, $U_{lj}$ are the mixing
matrix elements relating the flavor and mass neutrino eigenstates and
$m_{\nu j}$ denotes the Majorana mass of the $j$-th neutrino. 

Also, in the above expression we
have defined the leptonic current (the superscript $c$ denote the charge
conjugated spinor):
\be
L^{\alpha\beta}(p_1,p_2)=\bar{u}_l(p_1)\gamma^{\alpha}(1-\gamma_5)\gamma^{\beta}
u_l^c(p_2)
\label{18}
\ee
and the hadronic current 
\be
H_{\alpha\beta}(Q_i(q))= \sum_\eta
\bar{u}(p_B)\gamma_{\alpha}(f_{B\eta}+g_{B\eta}\gamma_5) 
\frac{\not Q_i+m_\eta}{Q_i^2-m_{\eta}^2}
\gamma_\beta (f_{A\eta}+g_{A\eta}\gamma_5)u(p_A)\ ,
\ee
where $f_{A\eta,\ B\eta}$ and $g_{A\eta,\ B\eta}$ are the vector and
axial-vector form factors for the single weak transitions of hyperons. The
subscript $\eta$ denotes the intermediate states that are allowed in each
specific transition (see Table II).

\begin{figure}
\includegraphics[width=15cm]{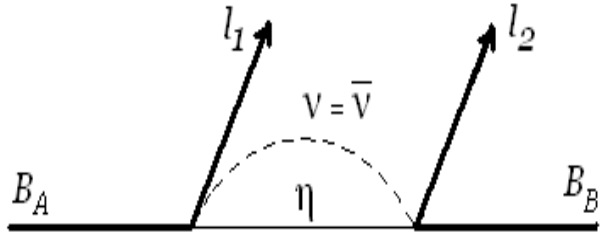}
\vspace{-10.0cm}
\caption{Feynman graph for $\Delta L=2$ hyperon decays. The virtual
state $\eta$ denotes an intermediate hyperon state.}
\end{figure}

  Since we will assume that our form factors are constants, the above
amplitude can be written in a more compact form as follows:
\be
{\cal M}_i=\sum_j m_{\nu j}U_{l j}^2\sum_\eta
\bar{u}(p_B)\gamma_{\nu}(f_{B\eta}+g_{B\eta}\gamma_5)
{\cal I}_i\gamma_{\mu}(f_{A\eta}+g_{A\eta}\gamma_5) 
u(p_A)L^{\nu \mu}(p_i,p_{3-i})\ ,
\ee
where we have introduced the loop integral:
\bea
{\cal I}_i &=& \int \frac{d^4q}{(2\pi)^4}
\frac{\not Q_i(q)+m_\eta}{(q^2-m_{\nu j}^2)(Q_i^2(q)-m_\eta^2)} \nonumber
 \\  
 &=& \frac{i}{(4\pi)^2}
\left[\frac{}{}(\not p_A-\not p_{3-i}){\cal
A}_\eta+m_\eta{\cal B}_\eta\right]\ .
\eea

  The integral ${\cal I}_i$ is logarithmically divergent. This divergence
can be cured in principle by taking into account the momentum dependence
of hyperon form factors, which are expected to fall for large
values of momentum transfer. Instead, we will
chose to introduce a momentum cut off $\Lambda$ \cite{Barbero:2002wm},
which can be related to the average
distance $d$ between quarks inside hyperons ($\Lambda \sim (2d)^{-1}
\approx 1$ GeV, for numerical purposes, as in our previous work
\cite{Barbero:2002wm}). Under this assumption, a
straightforward evaluation
of the functions ${\cal A}_{\eta},\ {\cal B}_{\eta}$ gives:
\bea
{\cal A}_\eta &=&{\cal C}_1\left({m_\eta^2}/{{\bar
m}_A^2},{\Lambda^2}/{{\bar m}_A^2}\right)-
{\cal D}_1\left({m_\eta^2}/{{\bar m}_A^2}\right),\nn \\
{\cal B}_\eta&=&{\cal C}_2\left({m_\eta^2}/{m_A^2},{\Lambda^2}/{{\bar
m}_A^2}\right)-
{\cal D}_2\left({m_\eta^2}/{{\bar m}_A^2}\right) \ ,
\eea
where we have defined $\bar{m}_A^2 = m_A^2+m_l^2$ and:
\bea
{\cal
C}_1(m,m')&=&-\frac{1}{2}(2+m)+\frac{1}{4}(1+m^2)\ln
\left(1+\frac{m}{m'}\right) +\frac{1}{2}\ln(m')\nn\\
&& +\ \frac{2m'-1+2mm'+m+m^2-m^3}{2\sqrt{4m'-(1-m)^2}}
\nn \\
&& \times 
\left[
\arctan\left(\frac{1-m}{\sqrt{4m'-(1-m)^2}}\right)+
\arctan\left(\frac{1+m}{\sqrt{4m'-(1-m)^2}}\right)\right]
,\nn\\
{\cal
C}_2(m,m')&=&-2+\frac{1}{2}(1+m)\ln
\left(1+\frac{m}{m'}\right) +\ln (m')   
+\frac{2m'-(1-m)^2}{\sqrt{4m'-(1-m)^2}}\nn\\
&& \times \left[
\arctan\left(\frac{1-m}{\sqrt{4m'-(1-m)^2}}\right)+
\arctan\left(\frac{1+m}{\sqrt{4m'-(1-m)^2}}\right)\right]
,\nn\\
{\cal D}_1(m)&=&-\frac{1}{2}(2+m)+\frac{1}{2} m^2\ln (m)+\frac{1}{2}(1-
m^2)\ln (1-m) +i\frac{\pi}{2}(1+m^2),\nn\\
{\cal D}_2(m)&=&-2+m\mbox{ln}(m)+(1- m)\mbox{ln}(1-m)
+i\pi(1+m)\ .
\eea
Since the integral in Eq. (7) (or ${\cal C}_{1,2}$) diverges
logarithmically, the dependence of our results on the cutoff $\Lambda$ is
not very sensitive.

\section{Results}

   In order to evaluate numerically the decay rates we need as input the
values of form factors. In Table II we show the numerical values of vector
and axial-vector form factors defined at zero momentum transfer taken
from a fit to hyperon semileptonic decays in the limit of SU(3)
\cite{Cabibbo:2003cu}. The subscript $A$ ($B$) refers to
the weak transition of initial (final) baryon state and $\eta$
denote the intermediate states that allowed to contribute in each case.
The values of  Cabibbo-Kobayashi-Maskawa matrix elements used in our
numerical evaluations are $|V_{ud}|=0.9740$ and
$|V_{us}|=0.2250$ \cite{pdg}.

\begin{table}
\begin{tabular}{|c|c|cccc|}
\hline
\ \ transition\ \  & \ $\ \eta\ $\  & \ \ \ $f_{A\eta}$\ \ \  &
\ \ \ $g_{A\eta}$ \ \ \ &
\ \ \ $f_{B\eta}$\ \ \ &\ \ \ $g_{B\eta}$
\\
\hline
$\Sigma^- \to \Sigma^+$ & $\Lambda$ & 0 & 0.656 & 0 & 0.656 \\
& $\Sigma^0$ & $\sqrt{2}$ & 0.655 & $\sqrt{2}$ & $-0.656$ \\
\hline
 & $n$ & $-1$ & \ 0.341\  & 1 & 1.2670 \\
$\Sigma^- \to p$ & $\Sigma^0$ & $\sqrt{2}$ & 0.655 & $-1/\sqrt{2}$ &
0.241 \\
 & $\Lambda$ & 0 & 0.656 & $-\sqrt{3/2}$ & $-0.895$ \\
\hline
 & $\Xi^0$ & $-1$ & 0.341 & 1 & 1.267 \\
$\Xi^- \to \Sigma^+$ & $\Sigma^0$ & $1/\sqrt{2}$ & 0.896 & $\sqrt{2}$ &
$-0.655$ \\
 & $\Lambda$ & $\sqrt{3/2}$ & $0.239$ & 0 & $0.656$ \\
\hline
$\Xi^- \to p$ & $\Sigma^0$ & $1/\sqrt{2}$ & 0.896 & $-1/\sqrt{2}$ & 0.241
\\
& $\Lambda$ & $\sqrt{3/2}$ & $0.239$ & $-\sqrt{3/2}$ & $-0.895$ \\
\hline
\end{tabular}
\caption{Vector ($f$) and axial-vector ($g$) form factors at zero
momentum transfer taken from Ref. \cite{Cabibbo:2003cu}. The second
column indicates the intermediate states that are allowed for each
transition.}  
 \end{table}

  Following the usual procedure, we can compute the decay rates from the
unpolarized probability obtained from the decay amplitudes given in
the previous section.  In the second column of Table III we show the decay
rates normalized to the square of the effective Majorana neutrino mass
which is defined as:
\be
\langle m_{l l} \rangle \equiv \sum_l m_{\nu j} U_{lj}^2  \ .
\ee
In the third column of Table III, we display the branching ratios for
each decay mode. Just for illustrative purposes we have used
$\langle m_{e e}
\rangle = 10$ eV and $\langle m_{\mu\mu} \rangle = 10$ MeV
 for the
effective Majorana masses of electron and muon neutrinos,
respectively. 

\begin{table}[h]
\bigskip
\begin{tabular}{c|c|c}
\hline
\hline
&\ \  ${\Gamma_{0\nu}}/{\langle m_{l l} \rangle^2}$ [sec$^{-1}/$MeV$^2$]\
\ &
$B(B_A \to B_B l^-l^-$) \\
\hline
$\Sigma^-\to \Sigma^+e^-e^-$&$1.000\x 10^{-15}$ & 1.48 $\times 10^{-35}$
\\
$\Sigma^-\to pe^-e^-$&$0.497\x 10^{-10}$ & 7.35 $\times 10^{-31}$\\
$\Sigma^-\to p\mu^-\mu^-$&$0.426\x 10^{-11}$ & 6.31 $\times 10^{-20}$\\ 
$\Xi^-\to \Sigma^+e^-e^-$&$0.841\x 10^{-13}$ & 1.38 $\times 10^{-33}$\\
$\Xi^-\to pe^-e^-$&$1.150\x 10^{-12}$ & 1.88 $\times 10^{-32}$ \\
$\Xi^-\to p\mu^-\mu^-$&$0.480\x 10^{-12}$ & 7.87 $\times 10^{-21}$\\
\hline\hline 
\end{tabular} 
\caption{Decay rates (normalized to the effective neutrino mass
$\langle m_{l l} \rangle^2$) and branching ratios for $\Delta
L=2$ hyperon decays. We use $\langle m_{e e} 
\rangle^2 =(10\ {\rm eV})^2$ and $\langle m_{\mu \mu} \rangle^2= (10\
{\rm MeV})^2$  to
evaluate the branching ratios.} 
\label{tab2}
\end{table}

   As we can expect \cite{Li:2007qx}, there is a considerable enhancement
in the rates of $\Delta S \not = 0$ transitions due mainly to a larger
phase available in such decays. Differences in the form factors for
different intermediate states play a less important role. On the other
hand, di-muon decays appear to have larger branching
ratios because the bounds on the effective Majorana mass of
muon neutrinos are rather poor at present. Conversely, if we use the
present experimental upper limit on the $\Xi^- \to p \mu^-\mu^-$ decay
\cite{Rajaram:2005bs} we derive the
following upper bound on the effective muon neutrino mass:
\be
\langle m_{\mu \mu} \rangle < 22\ {\rm TeV}\ .
\ee
Althought this bound is loosely compared to $\langle m_{\mu \mu}
\rangle \leq 0.04$ TeV obtained from $K^+ \to \pi^- \mu^+\mu^+$ decays
\cite{Littenberg:2000fg}, it is the first bound derived from $\Delta L=2$
hyperon decays. An improvement of 5 orders of magnitude on the
experimental upper limit of  the $\Xi^- \to p \mu^-\mu^-$ branching ratio 
would be required to produce a similar bound on $\langle m_{\mu \mu}
\rangle$. Note
also from Table III, that the $\Sigma^-\to p\mu^-\mu^-$ decay offers a
good chance to provide an upper limit on this effective Majorana mass
parameter although any experimental bound on this decay has been reported
up to now.

\section{Background: $\beta\beta$ decays with two neutrinos}

  Double beta decays with two neutrinos $B_A \to
B_Bl^-l^{'-}\bar{\nu}_l\bar{\nu}_{l'}$, which are allowed in the Standard
Model, can provide the main source of background for $\Delta L=2$ decays
of hyperons. Just for completeness, in this section we provide an estimate
of their branching fractions. 

   Following a model discussed in a previous work \cite{Barbero:2002wm},
we will assume the decays under consideration proceed through the decay
chain $B_A^- \to B^*l^-\bar{\nu}_l\to
B_B^+l^-l^{'-}\bar{\nu}_l\bar{\nu}_{l'}$, where $B^*$ is a neutral baryon
intermediate state. We further assume that the dominant contributions are
given by the $B^*$ states that belong to the same octet as $B_{A,B}$
\cite{Barbero:2002wm}. In this scheme, hyperon decays where intermediate
states $B^*$ can be on-shell simultaneously for the production and decay
subprocesses, will largely dominate the decay rate \cite{Barbero:2002wm}.
According with the convolution formula (see for example:
\cite{Calderon:2001qq}), the rates for the $2\bar{\nu}$ $\beta\beta$
decay processes are given to a good approximation by:
\be  
\Gamma(B_A^- \to B_B^+ l^-l^{'-}\bar{\nu}_l\bar{\nu}_{l'}) = \sum_{B^*}
\Gamma(B_A^- \to B^* l^-\bar{\nu}_l) \times B(B^* \to B_B^+
l^{'-}\bar{\nu}_{l'}) \ ,
\ee
where, for the decays of our interest, $B^*$ can be any of $n,\ \Lambda
,\ \Sigma^0$ and $\Xi_0$ real baryon states that are allowed by the
kinematics of the decay process.

\begin{table}
\bigskip
\begin{tabular}{c|c|c}
\hline
\hline
 & BR with all & BR with $\Sigma^0$ \\
Decay Mode & intermediate states& intermediate state \\
\hline\hline
$\Sigma^- \to \Sigma^+e^-e^-\bar{\nu}\bar{\nu}$ &$8.59\x 10^{-31}$ &
$8.59\x 10^{-31}$ \\
${}\ \to pe^-e^-\bar{\nu}\bar{\nu}$ & $1.02\x 10^{-3}$ & $2.85\x 10^{-23}$
\\
${}\ \to pe^-\mu^-\bar{\nu}\bar{\nu}$  &$4.5\x 10^{-4}$ & $1.23\x
10^{-23}$
\\
${}\ \to p\mu^-\mu^-\bar{\nu}\bar{\nu}$  &0 & 0 \\
$\Xi^- \to \Sigma^+e^-e^-\bar{\nu}\bar{\nu}$ & $6.59\x 10^{-14}$& $5.57\x 
10^{-25}$ \\
${} \ \ \to \Sigma^+e^-\mu^-\bar{\nu}\bar{\nu}$ & $1.20\x 10^{-15}$&
$6.78\x  10^{-27}$ \\
${} \ \to pe^-e^-\bar{\nu}\bar{\nu}$ & $4.68\x 10^{-7}$& $1.85\x 
10^{-17}$ \\
${} \to pe^-\mu^-\bar{\nu}\bar{\nu}$ & $3.80\x 10^{-7}$& $8.00\x 
10^{-18}$ \\
${} \to p\mu^-\mu^-\bar{\nu}\bar{\nu}$ & $5.49\x 10^{-8}$& $9.75\x 
10^{-20}$ \\
\hline 
\end{tabular}
\caption{Branching ratios for $2\bar{\nu}\ \beta\beta$ decays of
hyperons. Results in second column include all the allowed intermediate
baryon states, and in the third column we keep only contribution of the 
$\Sigma^0$ intermediate state (see text).} 
\label{tab4} \end{table}

  The results for the branching fractions are shown in Table IV, where we 
have used the results of ref. \cite{Garcia:1985xz} for the rates of single
beta hyperon decays. In the second column of Table IV, we display the
results obtained when all the on-shell $B^*$
intermediate states are allowed to contribute. Some branching fractions in
column 2 of Table IV appear to be surprisingly large, although they
correspond to an unrealistic situation. Indeed, in a real experiment,
contributions with  $n,\ \Lambda$ and $\Xi^0$ intermediate  states can be
discriminated and removed from data due to the large lifetimes of these
particles. 

A more realistic estimate of the branching ratios are given in the third
column of Table IV. These estimates include only the contribution of
$\Sigma^0$ as an intermediate state. Indeed, the $\Sigma^0$ can be
considered as an {\it irreducible} contribution given its very short
lifetime ($\tau_{\Sigma^0}= 7.4\x 10^{-20}$ sec.), which make appear the
two charged leptons as emitted from a common primary vertex.

   As a validation of our approximated formula in Eq. (12), we observe
that the branching  fraction for the $\Sigma^- \to
\Sigma^+e^-e^-\bar{\nu}\bar{\nu}$ transition ($8.59\x 10^{-31}$) is very
close to the result of the exact calculation ($1.36\x 10^{-30}$) obtained
in ref.
\cite{Barbero:2002wm}. 

\section{Conclusions}
 
   In this paper we have studied the $\Delta L=2$ transitions in hyperon
semileptonic decays. An explicit calculation of the branching ratios using
a model where loops are dominated by virtual baryons and Majorana
neutrinos shows that such decays are more suppressed than expectations
based on dimensional grounds \cite{Li:2007qx} and perhaps beyond the
scrutiny of present experiments. Using the present experimental bound on
the branching
ratio of $\Xi^- \to p\mu^-\mu^-$ decays, we get $\langle m_{\mu \mu}
\rangle \leq 22\ {\rm TeV}$ for the effective Majorana mass of the muon
neutrino. This bound is two orders of magnitude less restrictive than the
present bound on this parameter obtained from $K^+ \to \pi^-\mu^+\mu^+$
decays. Finally, it is interesting to note that, beyond any bound that can
be obtained on the efective Majorana masses, the observation of $\Delta
L=2$ hyperon transitions will signal the presence of new physics.

\acknowledgments

  G.L.C. acknowledges financial support from Conacyt (M\'exico).  C.B. and
A.M. are fellows of CONICET (Argentina) and acknowledge support under
grant PIP 06-6159.





\begin{thebibliography}{99}

\bibitem{oscillations}
Y.~Fukuda {\it et al.}  [Super-Kamiokande Collaboration],
  Phys.\ Rev.\ Lett.\  {\bf 81}, 1562 (1998)
  [arXiv:hep-ex/9807003];
 Phys.\ Rev.\ Lett.\  {\bf 82}, 2644 (1999)
  [arXiv:hep-ex/9812014]; 
 Phys.\ Rev.\ Lett.\  {\bf 86}, 5651 (2001)
  [arXiv:hep-ex/0103032]; 
K.~Eguchi {\it et al.}  [KamLAND Collaboration],
  Phys.\ Rev.\ Lett.\  {\bf 90}, 021802 (2003)
  [arXiv:hep-ex/0212021];  T.~Araki {\it et al.}  [KamLAND Collaboration],
  Phys.\ Rev.\ Lett.\  {\bf 94}, 081801 (2005)
  [arXiv:hep-ex/0406035]; Q.~R.~Ahmad {\it et al.}  [SNO Collaboration],
Phys.\ Rev.\ Lett.\  {\bf 87}, 071301 (2001)
  [arXiv:nucl-ex/0106015]; 
  Phys.\ Rev.\ Lett.\  {\bf 89}, 011301 (2002)
  [arXiv:nucl-ex/0204008].

\bibitem{Elliott:2002xe}
 See for example:  S.~R.~Elliott and P.~Vogel,
  Ann.\ Rev.\ Nucl.\ Part.\ Sci.\  {\bf 52}, 115 (2002)
  [arXiv:hep-ph/0202264]; S.~M.~Bilenky,
  J.\ Phys.\ G {\bf 32}, R127 (2006)
  [arXiv:hep-ph/0511227].

\bibitem{Barbero:2002wm}
  C.~Barbero, G.~L\'opez Castro and A.~Mariano,
  Phys.\ Lett.\  B {\bf 566}, 98 (2003)
  [arXiv:nucl-th/0212083].

\bibitem{pdg}
  W.~M.~Yao {\it et al.}  [Particle Data Group],
  J.\ Phys.\ G {\bf 33}, 1 (2006).

\bibitem{Littenberg:1991rd}
  L.~S.~Littenberg and R.~E.~Shrock,
  Phys.\ Rev.\  D {\bf 46}, 892 (1992).

\bibitem{Rajaram:2005bs}
  D.~Rajaram {\it et al.}  [HyperCP Collaboration],
  Phys.\ Rev.\ Lett.\  {\bf 94}, 181801 (2005)
  [arXiv:hep-ex/0505025].

\bibitem{Kodama:1995ia}
  K.~Kodama {\it et al.}  [E653 Collaboration],
  Phys.\ Lett.\  B {\bf 345}, 85 (1995).

\bibitem{Littenberg:2000fg}
  L.~S.~Littenberg and R.~Shrock,
  Phys.\ Lett.\  B {\bf 491}, 285 (2000)
  [arXiv:hep-ph/0005285].

\bibitem{Appel:2000tc}
  R.~Appel {\it et al.},
  Phys.\ Rev.\ Lett.\  {\bf 85}, 2877 (2000)
  [arXiv:hep-ex/0006003].

\bibitem{Li:2007qx}
  L.~F.~Li,
  arXiv:0706.2815 [hep-ph].

\bibitem{Cabibbo:2003cu}
  N.~Cabibbo, E.~C.~Swallow and R.~Winston,
  Ann.\ Rev.\ Nucl.\ Part.\ Sci.\  {\bf 53}, 39 (2003)
  [arXiv:hep-ph/0307298].

\bibitem{Calderon:2001qq}
  G.~Calder\'on and G.~L\'opez Castro,
  arXiv:hep-ph/0108088.

\bibitem{Garcia:1985xz}
  A.~Garcia, P.~Kielanowski and A.~Bohm (Ed.),
  Lect.\ Notes Phys.\  {\bf 222}, 1 (1985).

\end{thebibliography}
\end{document}